\documentclass[aps,prb,reprint,twocolumn,superscriptaddress,longbibliography]{revtex4-2}
\pdfoutput=1

\usepackage{graphicx}
\usepackage{amsmath,amsfonts,amssymb}
\usepackage{physics}
\usepackage{dsfont}
\usepackage{hyperref}
\usepackage{textcomp}
\usepackage{float}

\usepackage{cleveref}
\crefname{figure}{Figure}{Figures}
\crefname{table}{Table}{Tables}
\crefname{appendix}{Appendix}{Appendices}
\crefname{equation}{eq.}{eq.}

\begin{document}

\title{Nodal superconducting gap structure and topological surface states of UTe$_2$}

\author{Hans Christiansen$^1$, Max Geier$^2$, Brian M. Andersen$^1$, Andreas Kreisel}
\affiliation{Niels Bohr Institute, University of Copenhagen, DK-2200 Copenhagen, Denmark\\
$^2$Department of Physics, Massachusetts Institute of Technology, Cambridge, MA 02139, USA}

\date{\today}

\begin{abstract}
The heavy-fermion compound UTe$_2$ is a candidate for hosting intrinsic spin-triplet superconductivity. At present, however, the type of triplet Cooper pairing realized in UTe$_2$ remains unknown, which calls for further experimental and theoretical investigations. In this paper, we develop a microscopic minimal model for the superconducting phases of UTe$_2$ based on recent findings in the description of its low-energy normal state electronic properties. We apply the resulting theoretical model to extract the nodal gap properties of the allowed superconducting ground states, and determine their associated topological surface states on the experimentally relevant (0-11) cleave plane. We find that the Fermi surface of UTe$_2$ enforces additional point nodes in excess to the point nodes imposed by symmetry, which may reconcile several experiments seemingly in conflict with B$_{2u}$ or B$_{3u}$ pairing symmetries. Furthermore, we map out the in-gap Majorana surface-bound modes existing on the (0-11) surface, and discuss their potential for additional insight into the pairing structure of UTe$_2$.
\end{abstract}

\maketitle

\section{Introduction}

Since the discovery and subsequent understanding of spin-triplet ordered superfluidity in $^3$He, the condensed matter community has been on the hunt for equivalent exotic Cooper-paired phases of charged particles. Superconductivity with spin-triplet 
paired electrons is a rich phase of matter because the spin texture of the Cooper pairs can lead to new phenomena not present for spin-singlet ordered condensates. However, spin-triplet superconductivity has turned out to be seemingly rare in nature, particularly in the form arising from an intrinsic bulk superconducting instability of a paramagnetic metal. Identifying such materials and obtaining a quantitative description of their spin-triplet ordered superconducting state is important both for our basic understanding of the origin of unconventional pairing and for harvesting their potentially useful properties for future applications~\cite{Sarma2015,SatoFujimoto}. 

The heavy-fermion compound UTe$_2$ is currently a prime candidate for featuring an intrinsic spin-triplet superconducting phase in its ground state~\cite{Ran,Aoki_2022,Lewin_2023}.
Indications for this include that the superconducting state survives in large magnetic fields exceeding the Pauli limiting field, and the weak temperature dependence of the Knight shift upon entering the superconducting phase~\cite{Ran,Knebel2019,Nakamine2019,Aoki2019,Fujibayashi2022,Matsumura2023}.
UTe$_2$ is a body-centered orthorhombic material with D$_{2h}$ point group symmetry, allowing four symmetry-distinct spin-triplet order parameters in the case of strong spin-orbit coupling (SOC): A$_u$, B$_{1u}$, B$_{2u}$, and B$_{3u}$. Time-reversal symmetry can be broken below $T_c$ in the case of accidental near-degeneracies leading to superpositions of these states, as for example B$_{1u}$ + iB$_{2u}$. At present, the experimental status of the nature of superconductivity in UTe$_2$ remains controversial, with evidence in favor of all four allowed irreducible representations (irreps) and some also advocating for complex time-reversal symmetry-breaking non-unitary superpositions~\cite{Ishihara2023}. Thus, further experiments and theoretical modeling are needed to pinpoint the nature of the spin-triplet ordering of this material.

Several recent developments driven by improved sample quality appear to pave the way for significant progress in the understanding of superconductivity in UTe$_2$~\cite{Sakai2022}. The superconducting state is not reached by multiple thermodynamic transitions and time-reversal symmetry remains preserved in the ground state condensate~\cite{Ajeesh,Azari2023}. It appears that the earlier signatures of time-reversal symmetry breaking below $T_c$ originated from extrinsic sources~\cite{Clara1,Clara2,Iguchi2023,Andersen2024}. Thus, the superconducting state most likely resides within a single of the above-mentioned irreps~\cite{Theuss_2024}.
Recently, the normal state low-energy electronic structure and Fermi surface have been investigated in detail experimentally.
Quantum oscillations are consistent with corrugated cylindrical Fermi sheets~\cite{Eaton2024,Weinberger2024}. Evidence for a closed Fermi pocket~\cite{Fujimori2019,Miao2020,Broyles2023} is relatively weak, and hence we assume in this work the absence of closed Fermi pockets, which is important for the determination of the leading pairing instability and its associated topological properties~\cite{Ishizuka,Shishidou_2021,Henrik,Tei2023,Ohashi2024}. Given the extensive evidence for nodal quasiparticles from e.g. thermal conductivity~\cite{Metz2019,Hayes_thermal}, specific heat~\cite{Kittaka2020,Rosa2022} and penetration depth measurements~\cite{Ishihara2023}, the leading candidates for the superconducting order parameter appear to be B$_{2u}$ and B$_{3u}$.

Despite these important developments, many unsolved questions remain at present. Given that the properties highlighted above withstand further experimental scrutiny, which of the remaining two triplet phases, B$_{2u}$ or B$_{3u}$, are favored, i.e. does UTe$_2$ prefer to exhibit its point nodes on states made up of predominantly uranium or tellurium orbitals? Presumably, the answer to this question will eventually relate directly to the underlying mechanism driving this material into a triplet-ordered superconducting phase. Importantly, there are also many outstanding questions related to experiments {\it not} pointing directly to either B$_{2u}$ or B$_{3u}$ order~\cite{Suetsugu}. Examples come from the probed low-energy excitations extracted from microwave surface impedance and tunnel-diode oscillator experiments~\cite{Ishihara2023,Bae2021,Carlton-Jones}. Such probes are inconsistent with single pairs of symmetry-imposed point nodes along the high-symmetry directions of the crystal. This has led to discussions of multiple pairs of nodes off the high-symmetry axes and the influence of topological surface states on the screening currents~\cite{Ishihara2023,Carlton-Jones,WuWeyl}. Clearly, this calls for further experimental and theoretical studies of the superconducting phase of UTe$_2$, paving the way for a definitive determination of its ground state pairing structure, and thereby a description of both bulk and surface electronic properties. 

Motivated by these recent developments and remaining open questions, here we provide a microscopic minimal model for the superconducting phase of UTe$_2$ consistent with recent insight into its Fermi surface properties and the fundamental symmetries of the material. Often theoretical progress of superconductivity in heavy-fermion materials is impeded by the complexity of the correlated Kondo scattering in the normal state. Here we take a different approach that does not directly address the origin of triplet superconductivity, but rather builds on symmetry-allowed pairing terms given the Fermi surface extracted from recent quantum oscillation experiments~\cite{Eaton2024,Weinberger2024}. We focus on the distinguishing properties of the allowed irreps of the superconducting state. We find that in addition to the standard symmetry-imposed nodes for the B$_{u}$ phases, additional Fermi-surface imposed nodes must be present. This may be directly relevant for reconciling penetration depth measurements with non-chiral superconductivity~\cite{Ishihara2023}. We also extract the topological surface states attached to each irrep on the experimentally relevant (0-11) cleave plane, which yields additional possibilities to distinguish the pairing symmetry of UTe$_2$.

\section{Methodology}\label{sec:methods}

In this section, we first discuss the normal state band structure which is based on a slightly modified version of the density functional theory (DFT) calculations and subsequent four-band tight-binding model presented in Ref.~\onlinecite{Theuss_2024}. Second, we describe the derivation of the symmetry-allowed superconducting terms for both uranium (U) and tellurium (Te) sublattice sites. Finally, we outline the method used for extracting the surface Majorana bound states inside the superconducting gap.

\subsection{Model for the normal state}

\begin{figure}
    \includegraphics[width=0.99\linewidth]{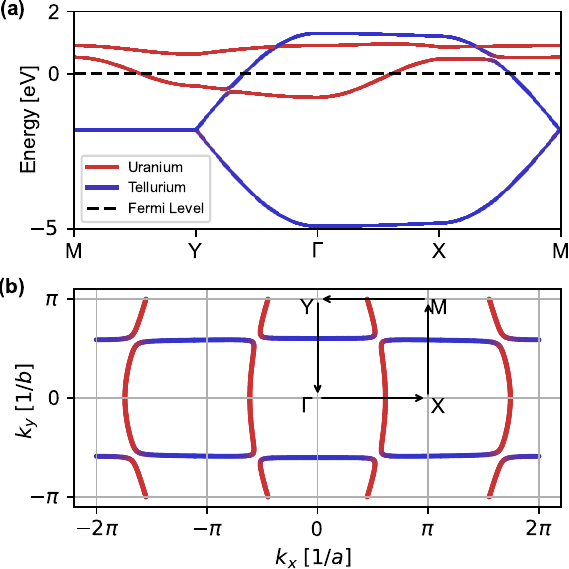}
    \caption{(a) Band structure of the applied four-band minimal model along the high-symmetry path indicated in panel (b) which displays the Fermi surface at $k_z=0$. Blue and red coloring indicate the Te and U content of the states, respectively. The full Fermi surface is shown in Fig.~\ref{fig:3dfs}.
    }
    \label{fig:bandstructure}
\end{figure}

\begin{figure}
    \includegraphics[width=0.99\linewidth]{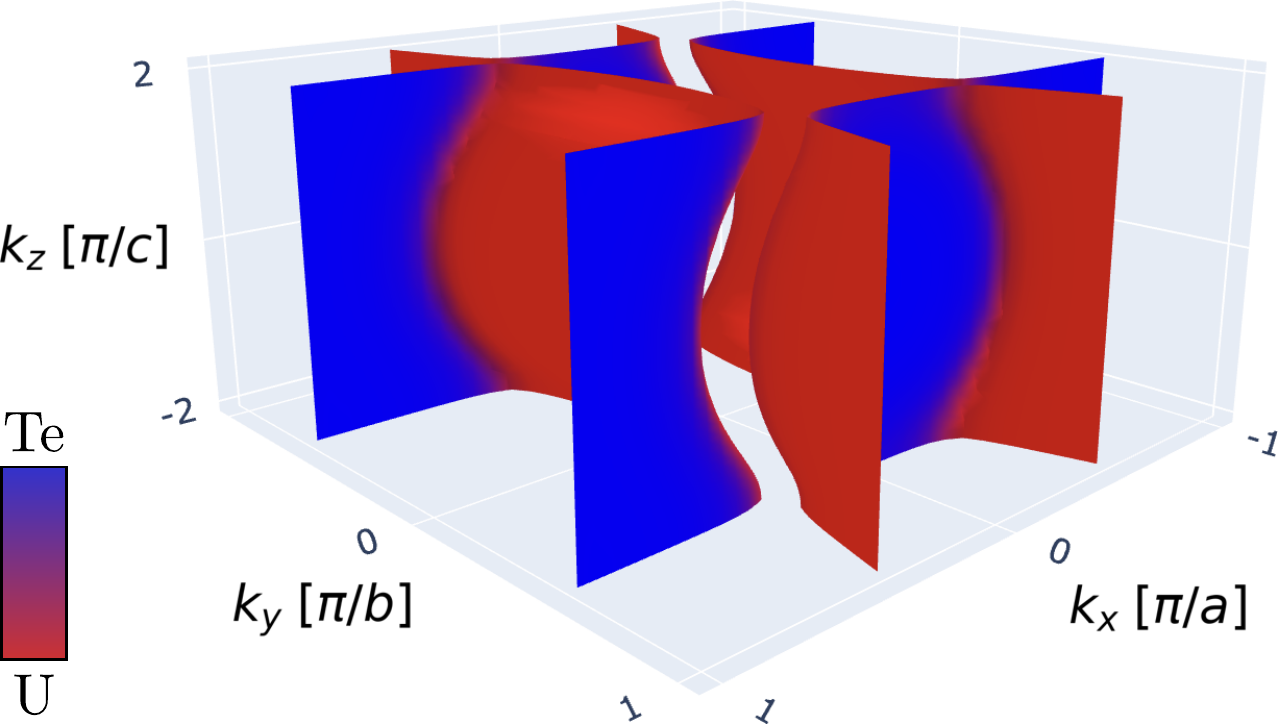}
    \caption{Three-dimensional plot of the Fermi surface applied in this work. Red (blue) color highlight the bands with dominant uranium (tellurium) orbital content.}
    \label{fig:3dfs}
\end{figure}

The starting point is a tight-binding model with few neighbor hoppings as used in Ref.~\onlinecite{Theuss_2024} to describe the electronic structure and Fermi surface, consistent with recent conclusions from quantum oscillation measurements~\cite{Eaton2024,Weinberger2024}. The Hamiltonian of the U degrees of freedom can be expanded in Pauli matrices $\rho_i$ as
\begin{align}
    H_{\mathrm{U}} =\phantom{-} \rho_0\;&[\mu_\mathrm U - 2t_\mathrm U \cos k_x - 2t_\mathrm {ch,U}\cos k_y] \notag \\
    -\rho_1\;&[ \Delta_\mathrm U + 2t_{\mathrm{U}}' \cos k_x + 2t_\mathrm {\rm ch,U}' \cos k_y \notag \\
    &\;- 4t_\mathrm {z,U} \cos(k_x/2)\cos(k_y/2)\cos (k_z/2)] \notag \\
    - \rho_2\;& 4 t_\mathrm {z,U}\cos(k_x/2)\cos(k_y/2)\sin(k_z/2) \notag    \\
    \equiv\phantom{-}\rho_0 \;&f_{\mathrm{U},0}(\mathbf{k}) + \rho_1\;f_{\mathrm{U},1}(\mathbf{k}) +\rho_2\;  f_{\mathrm{U},2}(\mathbf{k})\label{eq:HU},\
\end{align}
while for the Te degrees of freedom we have
\begin{align}
    H_{\mathrm{Te}} =\phantom{-} \rho_0\; &(\mu_{\mathrm{Te}} - 2t_{ch,\mathrm{Te}}\cos k_x) \notag \\
    -\rho_1  \; & [\Delta_{\mathrm{Te}} - t_{\mathrm{Te}}\cos k_y  \notag\\
    &-2t_{z,\mathrm{Te}}\cos(k_x/2)\cos(k_y/2)\cos(k_z/2)]\notag \\
    -\rho_2\; & t_{\mathrm{Te}}\sin k_y  \notag\\
    \equiv\phantom{-}\rho_0 \;& f_{\mathrm{Te},0}(\mathbf{k}) + \rho_1 \;f_{\mathrm{Te},1}(\mathbf{k})+\rho_2\; f_{\mathrm{Te},2}(\mathbf{k}) \label{eq:HTe}.\
\end{align}
Here, $\rho_i$ denote Pauli matrices in sublattice space for both the U and Te sectors. 
Together with a hybridization term of the form
\begin{equation}
    H_{\mathrm{Te}-\mathrm U} =  \delta (\rho_0+\rho_1) ,\label{eq_hyb}
\end{equation}
the full normal state Hamiltonian is given by
\begin{equation}
    H_N(\mathbf k) = \begin{pmatrix}
        H_{\mathrm{U}} && H_{\mathrm{U}-\mathrm{Te}} \\
        H_{\mathrm{U}-\mathrm{Te}}^\dag && H_{\mathrm{Te}}
    \end{pmatrix} \label{eq_H_normal}.\
\end{equation}

The allowed form of the hybridization term can be deduced from the representations of the $\mathrm{D}_{2h}$ mirror symmetries given in Tab. \ref{tab:symreps}. Under $\mathcal M_z$ symmetry, the hybridization terms transform as
\begin{equation}
    \begin{pmatrix}
        && H_{\mathrm{U}-\mathrm{Te}} \\
        H_{\mathrm{U}-\mathrm{Te}}^\dag &&
    \end{pmatrix}
    \mapsto
    \begin{pmatrix}
        && \rho_1\; H_{\mathrm{U}-\mathrm{Te}} \\
        H_{\mathrm{U}-\mathrm{Te}}^\dag\;\rho_1 &&
    \end{pmatrix},
\end{equation}
from which we read off that Eq.~(\ref{eq_hyb}) is the only constant-in-momentum hybridization term that respects $\mathcal M_z$. Similarly, this term remains invariant under $\mathcal M_y$ symmetry, under which the hybridization transforms as $H_{\mathrm{U-Te}} \mapsto H_{\mathrm{U-Te}}\rho_1$.

\begin{table}[t]
    \centering
    \begin{tabular}{|c|c|l|l|l|} \hline
         &  $\mathcal M_x$ & $\mathcal M_y$&$\mathcal M_z$ &$\mathcal I$\\ \hline
         Uranium&
     $\rho_0$ & $\rho_0$ &  $\rho_1$ & $\rho_1$ \\ \hline
 Tellurium& $\rho_0$& $\rho_1$ &  $\rho_0$ & $\rho_1$ \\ \hline
 \end{tabular}
    \caption{Representations of the mirror and inversion operations of the $D_{2h}$ point group on the blocks of the Hamiltonian in Eq.~(\ref{eq_H_normal}).}
    \label{tab:symreps}
\end{table}

\begin{table*}
    \centering
    \begin{tabular}{|c|c|c|c|c|c|c|c|c|c|c|c|c|c|} \hline
        Parameters [eV] & $\mu_U$ & $\Delta_U$ & $t_U$ & $t_U'$ & $t_{ch,U}$ & $t_{ch,U}'$ & $t_{z,U}$ & $\mu_{\mathrm{Te}}$ & $\Delta_{\mathrm{Te}}$ & $t_{\mathrm{Te}}$ & $t_{ch,\mathrm{Te}}$ & $t_{z,\mathrm{Te}}$ & $\delta$  \\ \hline
        Used in \cite{Theuss_2024} & 0.4 & 0.4 & 0.15 & 0.08 & 0.01 & 0 & -0.03 & -1.8 & -1.5 & -1.5 & 0 & -0.05 & 0.09 \\ \hline
        This work & 0.4 & 0.35 & 0.15 & 0.08 & 0.01 & 0 & 0.08 & -1.8 & -1.5 & -1.5 & 0 & -0.05 & 0.045
     \\ \hline\end{tabular}
    \caption{Parameters used in the normal-state tight-binding Hamiltonian.}
    \label{tab:ute2parameters}
\end{table*}

The parameters used for the normal state tight-binding band structure are displayed in Tab.~\ref{tab:ute2parameters}. The minor changes of the parameters compared to Ref.~\onlinecite{Theuss_2024} are motivated by recent low-energy tunneling measurements~\cite{SeamusQPI}. Distinct from Ref.~\onlinecite{Theuss_2024} we write the hybridization term in sublattice space. The main effect of these modified parameters is to produce slightly more dispersive Fermi surface sheets with dominant U content along $k_z$.

We stress that the normal state low-energy electronic structure of UTe$_2$ remains controversial at present~\cite{Broyles2023,Eaton2024,AokiQO2022,Eo2022}. In particular, the role of Kondo correlations and their associated temperature-dependent effects on the Fermi surface are topics of intense research~\cite{Aoki2019,Xu2019,Shick2019,IshizukaPRB2021,Choi2024,LiuDMFT}. There are also experimental signatures from photoemission of spectral weight around the Z point in the Brillouin zone (BZ)~\cite{Fujimori2019,Miao2020}. The theoretical framework of the present work does not include normal state effects beyond the tight-binding model presented above. A Fermi surface consisting of cylindrical 2D sheets is consistent with several first-principles calculations able to include strong electronic correlations~\cite{Aoki2019,Xu2019,Henrik}. In the following, under the assumption that the electronic band structure presented in Figs.~\ref{fig:bandstructure} and \ref{fig:3dfs} faithfully represents the topology of the Fermi surface in UTe$_2$, we deduce the consequences for the superconducting gap structure and surface states on the experimentally relevant (0-11) cleave plane.

\subsection{Model for the superconducting  state}
As a standard approach, we now formulate the superconducting Hamiltonian in a Bogoliubov-de Gennes (BdG) framework where the Hamiltonian is trivially expanded to (pseudo-)spin space and additionally to particle-hole space such that it reads
\begin{equation}\label{eq:bdgham}
    H(\mathbf k) = \begin{pmatrix}
        H_N(\mathbf k)\sigma_0 && \Delta(\mathbf k) \\
        \Delta^\dag(\mathbf k) && -H_N^*(-\mathbf k)\sigma_0
    \end{pmatrix} ,\
\end{equation}
where the order parameter matrix is defined by
\begin{equation}
 \Delta(\mathbf k)=(\vec d(\mathbf k) \cdot \vec{\sigma})i\sigma_y,
\end{equation}
with the $\vec d$-vector describing generic triplet superconductivity; each component is a matrix in sublattice space $\rho_i$ and the space of U and Te as in Eq.~(\ref{eq_H_normal}). The symmetry of the system determines the possible pairing symmetry of the superconducting order parameter. As a result, the system will most likely condense into a superconducting phase characterized by an order parameter that transforms as one of the irreps of the symmetry group. Based on the current experimental findings on the superconducting state of UTe$_2$, i.e. restricting to triplet superconductivity, the $\vec d$-vector should transform as one of the odd-parity irreps of the $D_{2h}$ point group. In the limit of strong SOC, the four irreps exhibit the components as given in Table~\ref{tab:dvector}.

\begin{table}[tb]
        \begin{tabular}{|l|l|}
        \hline
        Irreps & $d$ symmetry in band space \\
        \hline
        $A_u$&$(AX,BY,CZ)$ \\
         $B_{1u}$& $(AY,BX,CXYZ)$ \\
         $B_{2u}$& $(AZ,BXYZ,CX)$ \\
         $B_{3u}$& $(AXYZ,BZ,CY)$ \\
        \hline
     \end{tabular}
     \caption{Possible forms of the $\vec d$-vector for each symmetry-allowed odd-parity irrep of the $D_{2h}$ point group, assuming strong SOC (i.e. pseudospin and momentum transform together). Here, $X,Y,Z$ denote any function odd in $k_x,k_y,k_z$, respectively~\cite{hillier2012nonunitary}.}
    \label{tab:dvector}
\end{table}

\begin{table*}[]
    \begin{minipage}[t]{0.48\textwidth}
        \begin{tabular}{|l|l|l|l|}
            \hline
                     &   U sector&Te sector &U-Te sector\\
                     \hline
                     $X$& 
                 $f_x \rho_{1/0}$, $f_{xz}\rho_2$&$f_x\rho_{1/0}$, $f_{xy} \rho_2$&$f_x(\rho_0+\rho_1)$, $f_{xyz}(\rho_0-\rho_1)$\\ \hline
             $Y$& $f_y\rho_{1/0}$, $f_{yz}\rho_2$&$f_y\rho_{1/0}$, $f_1\rho_2$&$f_y(\rho_0+\rho_1)$, $f_z(\rho_0-\rho_1)$\\ \hline
             $Z$& $f_z \rho_{1/0}$, $f_1\rho_2$&$f_z\rho_{1/0} $, $f_{yz}\rho_2$&$f_z(\rho_0+\rho_1)$, $f_{y}(\rho_0-\rho_1)$\\
             \hline
             $XYZ$ & $f_{xyz}\rho_{1/0}$, $f_{xy}\rho_2$&$f_{xyz}\rho_{1/0}$, $f_{xz}\rho_2$&$f_{xyz}(\rho_0+\rho_1)$, $f_x(\rho_0-\rho_1)$\\
             \hline 
        \end{tabular}
        \caption{Symmetry-allowed superconducting terms in sublattice space that transform according to the leftmost column. Here, $X$ is shorthand for a momentum-dependent function that transforms like $k_x$ under all point group symmetries of D$_{2h}$ and $f_i$ ($f_{xyz}$) are lattice harmonics that transform as $k_i$ ($k_xk_yk_z$). $f_1$ transforms trivially under $D_{2h}$.}
        \label{tab:basisfunctions}
    \end{minipage}
    \hfill
    \begin{minipage}[t]{0.48\textwidth}
        \begin{tabular}{|l|l|l|}
            \hline
                     &   $i=1$&$i=2$\\
                     \hline
                     $f_{x,i}$& 
                 $\sin(k_x/2)\cos(k_y/2)\cos(k_z/2)$&$\sin k_x$\\ \hline
             $f_{y,i}$& $\cos(k_x/2)\sin(k_y/2)\cos(k_z/2)$&$\sin k_y$\\ \hline
             $f_{z,i}$& $\cos(k_x/2)\cos(k_y/2)\sin(k_z/2)$&$\sin k_z$\\\hline
 $f_{xy}$& $\sin(k_x/2)\sin(k_y/2)\cos(k_z/2)$&$-$\\\hline
 $f_{xz}$& $\sin(k_x/2)\cos(k_y/2)\sin(k_z/2)$&$-$\\\hline
 $f_{yz}$& $\cos(k_x/2)\sin(k_y/2)\sin(k_z/2)$&$-$\\\hline
             
             $f_{xyz}$& $\sin(k_x/2)\sin(k_y/2)\sin(k_z/2)$&$-$\\ \hline 
        \end{tabular}
        \caption{Lattice harmonics $f_i=f_i(\mathbf k)$ corresponding to nearest and next-nearest neighbor pairing in the superconducting order parameter.}
        \label{tab:latticeharmonics}
    \end{minipage}
\end{table*}

Next, we classify all allowed terms in sublattice space, giving rise to the correct superconducting order in band space~\cite{Shishidou_2021}. To do so, we first consider the tellurium and uranium sectors separately by setting the hybridization $\delta=0$. In this case, both the uranium and tellurium normal state Hamiltonians have the form $H_\alpha=\rho_0f_{\alpha,0}(\mathbf{k})+\rho_1f_{\alpha,1}(\mathbf{k})+\rho_2 f_{\alpha,2}(\mathbf{k})$, $\alpha= \mathrm{U}/\mathrm{Te}$ and can be diagonalized analytically. We transform to their respective band basis where $\Tilde{H}_\alpha =\mathrm{diag}\left(f_{\alpha,0}\pm\sqrt{f_{\alpha,1}^2+f_{\alpha,2}^2}\right)$. This can be performed by the following unitary transformation
\begin{equation}
    u_\alpha (\mathbf{k}) = 
    \frac{1}{\sqrt{2}}\begin{pmatrix}
        -1 && \frac{f_{\alpha,1}(\mathbf{k}) -if_{\alpha,2}(\mathbf{k})}{\sqrt{f_{\alpha,1}(\mathbf{k})^2+f_{\alpha,2}(\mathbf{k})^2}} \\
        \frac{f_{\alpha,1}(\mathbf{k}) + if_{\alpha,2}(\mathbf{k})}{\sqrt{f_{\alpha,1}(\mathbf{k})^2+f_{\alpha,2}(\mathbf{k})^2}} && 1
    \end{pmatrix},\
\end{equation}
which is also Hermitian. We now transform each $\alpha$ block of the BdG Hamiltonian Eq.~(\ref{eq:bdgham}), with the unitary matrix
\begin{equation}
    U(\mathbf{k}) = \begin{pmatrix}
        u(\mathbf{k}) && 0 \\
        0 && u^*(-\mathbf{k})
    \end{pmatrix} ,\
\end{equation}
where we drop the $\alpha$ label here and in the following discussion. Using that $u^*(-\mathbf{k})=u(\mathbf{k})$, we find that the sublattice matrices in both the normal and anomalous sectors of the BdG Hamiltonian have the form $\Tilde{\rho_i}(\mathbf{k}) \equiv  u^\dag(\mathbf{k}) \rho_i u(\mathbf{k})$ given by 

\begin{equation}
    \Tilde{\rho}_1 = \begin{pmatrix}
        \frac{-f_1}{\sqrt{f_1^2+f_2^2}} & \frac{-if_2}{f_1+if_2} \\
        \frac{if_2}{f_1-if_2} & \frac{f_1}{\sqrt{f_1^2+f_2^2}}
    \end{pmatrix},\
\end{equation}

\begin{equation}
    \Tilde{\rho}_2 = \begin{pmatrix}
        \frac{-f_2}{\sqrt{f_1^2+f_2^2}} & \frac{if_1}{f_1+if_2} \\
        \frac{-if_1}{f_1-if_2} & \frac{f_2}{\sqrt{f_1^2+f_2^2}}
    \end{pmatrix},\
\end{equation}

\begin{equation}
    \Tilde{\rho}_3 = -\begin{pmatrix}
        0 && \frac{f_1-if_2}{\sqrt{f_1^2+f_2^2}} \\
        \frac{f_1+if_2}{\sqrt{f_1^2+f_2^2}} && 0
    \end{pmatrix} ,\
\end{equation}
where we suppress the $\mathbf{k}$ dependence for clarity. From the diagonal elements of these matrices we can read off that $\rho_1$ (and also $\rho_0$) transform trivially while $\rho_2$ transforms like $f_2(\mathbf{k})$, meaning that it transforms as $Z$ in the uranium sector and $Y$ in the tellurium sector, as seen from Eqs.~(\ref{eq:HU}) and (\ref{eq:HTe}). From this knowledge of the transformations of the sublattice matrices, we can combine with an appropriate spin vector and momentum-dependent function to obtain all symmetry-allowed terms for each irrep, as summarized in Tab.~\ref{tab:basisfunctions}. 

In the presence of finite hybridization, $\delta\neq 0$, the band transformation becomes more complicated. However, by inspecting the representations of the generators of the $D_{2h}$ point group in Tab. \ref{tab:symreps}, one can verify that the superconducting terms written in Tab. \ref{tab:basisfunctions} still transform correctly under all symmetry transformations. Finally, this also allows for the extension to superconducting terms in the hybridized U-Te sector written in the last column of Tab. \ref{tab:basisfunctions}, using the fact that $\rho_0+\rho_1$ is even while $\rho_0-\rho_1$ is odd under $\mathcal M_z$ and $\mathcal M_y$ symmetries.

\subsection{Accessing the surface states}\label{sec:surface states}

In the following, we analyze both the bulk and surface electronic properties of the superconducting phases of UTe$_2$. In order to capture the surface states, we assume periodic boundary conditions in the two directions that lie in the surface plane, such that the parallel momentum $\mathbf k^{\parallel}$ remains a good quantum number. In the third direction (normal to the surface), the open boundary conditions break the translation invariance such that the full Hamiltonian can be written as
\begin{align}\label{eq:tridiag1}
    &H(\mathbf k^{\parallel})  =     \begin{pmatrix}
    H_0 & V  & 0  & \hdots &  \\
    V^\dag & H_0 & V & 0  & \hdots  \\
    0 & V^\dag & H_0 & \ddots &   \\
    \vdots & 0 & \ddots & \ddots & V  \\
    & \vdots & & V^\dag & H_0
    \end{pmatrix}\;,
\end{align}
where $H_0$ describes the terms in the (BdG) Hamiltonian within a (suitably) chosen layer and $V$ is the coupling between nearest neighbor planes.
The retarded Green's function is given by 
\begin{align}\label{eq:Gtridiag}
    G(\omega,\mathbf k^{\parallel})  &=
    \frac{1}{\omega + i\eta - H(\mathbf k^{\parallel})}  \notag \\ 
    &=\begin{pmatrix}
    G_s & \hdots  &  &  &  \\
    \vdots & \ddots &  &  &   \\
     & & G_{b} &  &   \\
     & & & \ddots &  \vdots \\
    & & & \hdots & \tilde{G_{s}} 
    \end{pmatrix}  ,\
\end{align}
where the highlighted blocks denote the inter-layer component on the two surfaces and in the bulk. Given the Hamiltonian form in Eq.~(\ref{eq:tridiag1}), one can use the numerical method described in Ref.~\onlinecite{sancho_highly_1985} to efficiently obtain $(G_s,\tilde{G_s},G_b)$. We refer to Appendix~\ref{app:B} for further details on how to extract these Green's functions. Throughout this paper, we use $\eta=\Delta_0/300$ with $\Delta_0=10  \: \mathrm{meV}$ where $\Delta_0$ is the overall prefactor of the superconducting terms. From the surface and bulk Green's functions, the observables of interest can be computed. For example, the electronic spectral density within a specific sector can be obtained by  
\begin{equation}
    A_{s/b,P}(\omega,\mathbf k^{\parallel}) = \Tr{ P_e\otimes P(\mathbf k^\parallel)
    G_{s/b}(\mathbf k^\parallel,\omega)} ,\
\end{equation}
where $P_e = \mathrm{diag}(1,0)$ is the projector in the Nambu space onto the electronic sector, and $P(\mathbf k^\parallel)$ denotes the normal state projector onto the relevant sector, e.g. $P(\mathbf k^\parallel)=\sigma_i$ for the electronic spin polarization. 

Finally, to properly describe the surface states on the experimentally relevant (0-11) cleave plane, we need to define a coordinate system that respects the symmetries of that specific surface. In particular, this requires introducing a new set of lattice vectors such that two of them leave the (0-11) surface invariant in addition to the crystal lattice. From this we define a surface parallel momentum $\mathbf{k}^\parallel =k_x\vec m_x + k_{c^*} \vec m_{c^*}$ in terms of $\vec m_x = \hat e_x/a$ and $\vec m_{c^*} = \hat e_y /b + \hat e_z/c$, where $a$, $b$ and $c$ are the lattice constants and $\hat e_i$, $i=x,y,z$, are the Cartesian unit vectors. Due to the diagonal nature of the surface, the reciprocal lattice vectors are not proportional to the unit vectors in this coordinate system. Instead, they inherit the body-centered orthorhombic periodicity of the bulk unit cell such that the reciprocal lattice vectors have the forms $(2\pi,0)$ and $(\pi,2\pi)$, see Appendix~\ref{app:A} for more details. In Fig.~\ref{fig:surfacebz} we show the resulting surface Brillouin zone projected onto the (0-11) surface and the reciprocal lattice vectors showing the periodicity of the BZ. We also plot the uranium part of the Fermi surface while leaving the tellurium sector out because it gives an almost constant background weight as a result of the particular orientation of the (0-11) plane and the rather flat dispersion of the Te bands.

\begin{figure}
    \centering
    \includegraphics[width=\linewidth]{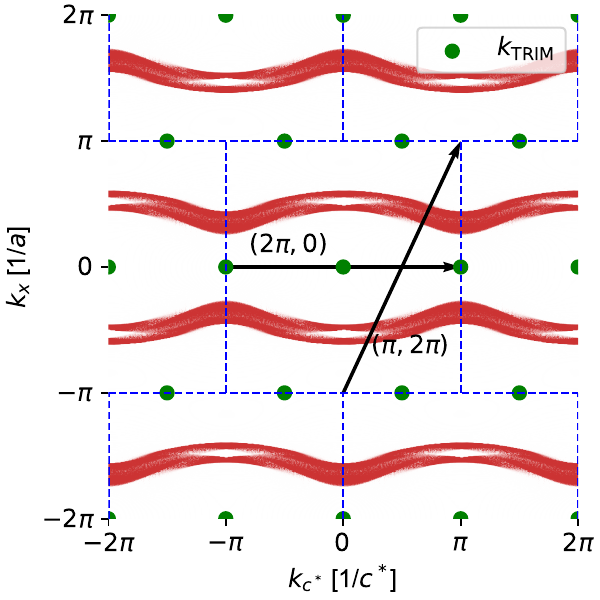}
    \caption{Surface Brillouin zone of the (0-11) surface with the reciprocal lattice vectors $(2\pi,0)$ and $(\pi,2\pi)$, as shown in Appendix~\ref{app:A}. The green dots show the time-reversal invariant momenta (TRIM) corresponding to this periodicity. The red bands display the uranium part of the projected spectral function at $\omega=0$.}
    \label{fig:surfacebz}
\end{figure}

\section{Results}\label{sec:results}

\subsection{Bulk superconducting gap structure} \label{sec:gapnodalstructure}

To obtain the symmetry-allowed terms in the BdG Hamiltonian describing each of the superconducting phases, we include all combinations with spin components according to Tab.~\ref{tab:dvector} and momentum symmetries according to Tab.~\ref{tab:basisfunctions}. We include momentum functions up to, and including, next-nearest neighbor coupling as seen in Tab.~\ref{tab:latticeharmonics}. 

\begin{figure*}[tb]
    \centering
    \includegraphics[width=\linewidth]{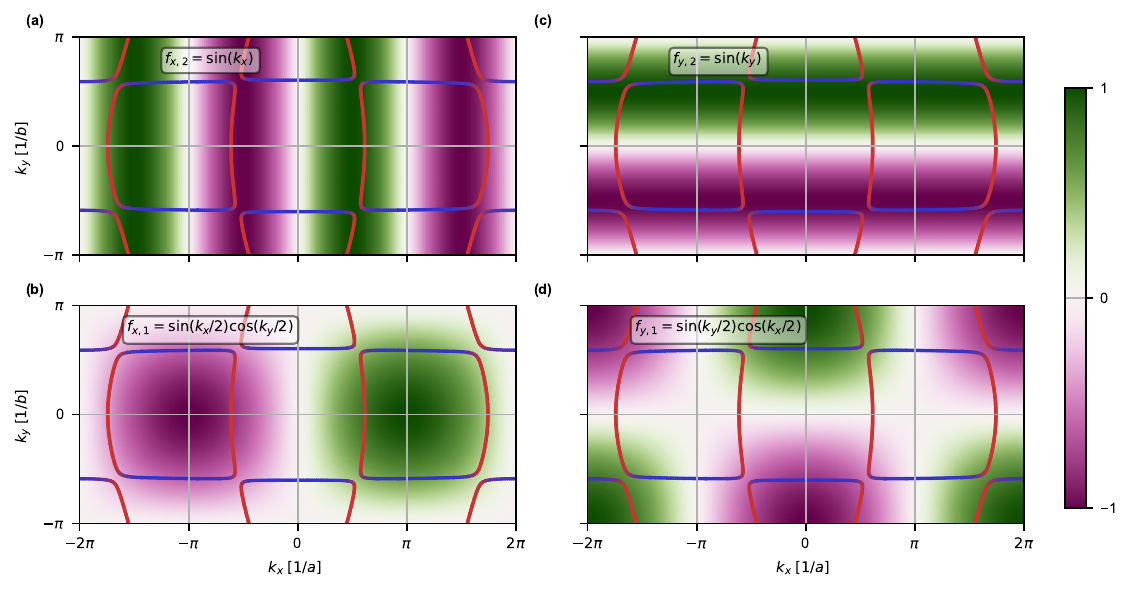}
    \caption{Fermi surface at $k_z=0$ along with the possible momentum functions $f_{x,1}$ and $f_{y,1}$ describing nearest neighbor pairing and $f_{x,2}$, $f_{y,2}$ describing next-nearest neighbor pairing for the B$_{2u}$ (left) and B$_{3u}$ (right) phases. Note that B$_{2u}$ (B$_{3u}$) is odd (even) under $k_x\rightarrow-k_x$ which is a distinction that is important for the distinct topological properties of these two phases on the experimentally relevant  (0-11) surface. }
    \label{fig:2dfs_basisfunctions}
\end{figure*}

The main qualitative difference between the different superconducting phases comes from the resulting nodal structure. Inspecting Tab.~\ref{tab:dvector} we see, as expected, that the B$_{2u}$ and B$_{3u}$ phases have symmetry-imposed nodes along the $k_y$ and $k_x$ axes, respectively. By contrast, the A$_u$ phase is fully gapped independently of the Fermi surface topology, while the B$_{1u}$ phase is fully gapped in the present case because the Fermi surface does not cross the nodal $k_z$ axis. Experiments indicate nodal quasiparticles~\cite{Metz2019,Hayes_thermal,Rosa2022,Ishihara2023} and therefore we consider the B$_{2u}$ and B$_{3u}$ phases in the following.

Aside from the symmetry-imposed nodes discussed above, both the B$_{2u}$ and the B$_{3u}$ phases exhibit additional nodes that arise due to a combination of the symmetry of the order parameter and the body-centered orthorhombic periodicity of the BZ, as explained in detail below. The precise locations of these additional nodes are not imposed by symmetry. We do stress, however, that they are true nodes and not artifacts of including only the lowest-order pairing terms, i.e. including longer-range pairing will not gap them out.

\begin{table}[]
    \centering
    \begin{tabular}{|c|c|c|c|c|} \hline 
         Irrep/$g$&  $\mathcal M_x$ & $\mathcal M_y$&$ \mathcal M_z$ &$\mathcal I$\\ \hline 
         A$_{u}$& 
     $-1$& $-1$& $-1$&$-1$\\ \hline 
 B$_{1u}$& $1$& $1$& $-1$&$-1$\\ \hline 
 B$_{2u}$& $1$& $-1$& $1$&$-1$\\ \hline 
 B$_{3u}$& $-1$& $1$& $1$&$-1$\\ \hline\end{tabular}
    \caption{Character tables for generators $g$ of the D$_{2h}$ point group.}
    \label{tab:chractertable}
\end{table}

To demonstrate this, consider the superconducting terms in the B$_{2u}$ phase in the $k_z=0$ plane. Here, the $\vec d$-vector only has one component $\vec d \sim (0,0,X)$ so it is the relevant plane to consider for the nodal structure of this order parameter. Since only one component is relevant at $k_z=0$, we consider the two types of momentum dependencies transforming as $X$ in the first row of Tab.~\ref{tab:basisfunctions}. Assuming momentarily that the superconducting phase could avoid all but the symmetry-imposed nodes, this would require a dominant $f_{x,2}=\sin k_x$ pairing in the uranium sector since the nodal lines of $f_{x,2}$ have no overlap with the uranium Fermi surface, as seen in Fig.~\ref{fig:2dfs_basisfunctions}(a). A similar argument requires a dominant $f_{x,1}=\sin(k_x/2)\cos(k_y/2)\cos(k_z/2)$ pairing in the tellurium sector, as evident from Fig.~\ref{fig:2dfs_basisfunctions}(b). In the absence of any hybridization between the tellurium and uranium bands, this gap structure would result in a superconducting phase with only the symmetry-imposed nodes. However, at half of the band crossing points, the order parameter necessarily exhibits a different sign in the uranium and tellurium sectors, as seen by comparing Figs.~\ref{fig:2dfs_basisfunctions}(a,b) where four such points can be located. As a result of this sign change, any finite hybridization induces nodes at the avoided level crossing. In order to see this, consider a point in momentum space close to the band-crossing region. In that case the BdG Hamiltonian takes the simple form
\begin{align}\label{eq:additionalnodes}
    &H =     \begin{pmatrix}
    \xi & \delta  & \Delta_{\rm U}  & 0   \\
    \delta & \xi & 0 & \Delta_{\rm Te}   \\
    \Delta_{\rm U} & 0 & -\xi & -\delta   \\
    0 & \Delta_{\rm Te} & -\delta  & -\xi\\
    \end{pmatrix},  
\end{align}
where $\xi_{\rm U}=\xi_{\rm Te}=\xi$ at the band crossing point, with $\xi_{\rm U}$ and $\xi_{\rm Te}$ denoting the U and Te bands in the absence of hybridization $\delta$. We are interested in regions of momentum space where the band crossing points exhibit opposite signs of the order parameter. Thus, let us assume simply $\Delta_{\rm U}=-\Delta_{\rm Te}=\Delta$. Transforming to a basis where the hybridized bands are diagonal, we obtain $\Delta \tau_z \rightarrow \Delta \tau_x$. Therefore, the hybridized bands necessarily feature a node near the band crossing point, which is in addition to the symmetry-imposed nodes along the high-symmetry $k_y$ axis. For a more general set of parameters, the nodes move to other positions of the Fermi surface as elaborated further below. We find that these nodes can only be removed by inter-band pairing terms which are unlikely to play a role for the tiny energy scales of the superconducting order relevant for UTe$_2$.

The same reasoning can be applied to the case of B$_{3u}$ pairing. In that case, the $\vec d$-vector has the form $\vec d \sim (0,0,Y)$ in the $k_z=0$ plane, so the relevant pairing functions are $f_{y,1}$ and $f_{y,2}$ as shown in Figs.~\ref{fig:2dfs_basisfunctions}(c,d). 

\begin{figure}[tb]
    \centering
    \includegraphics[width=1\linewidth]{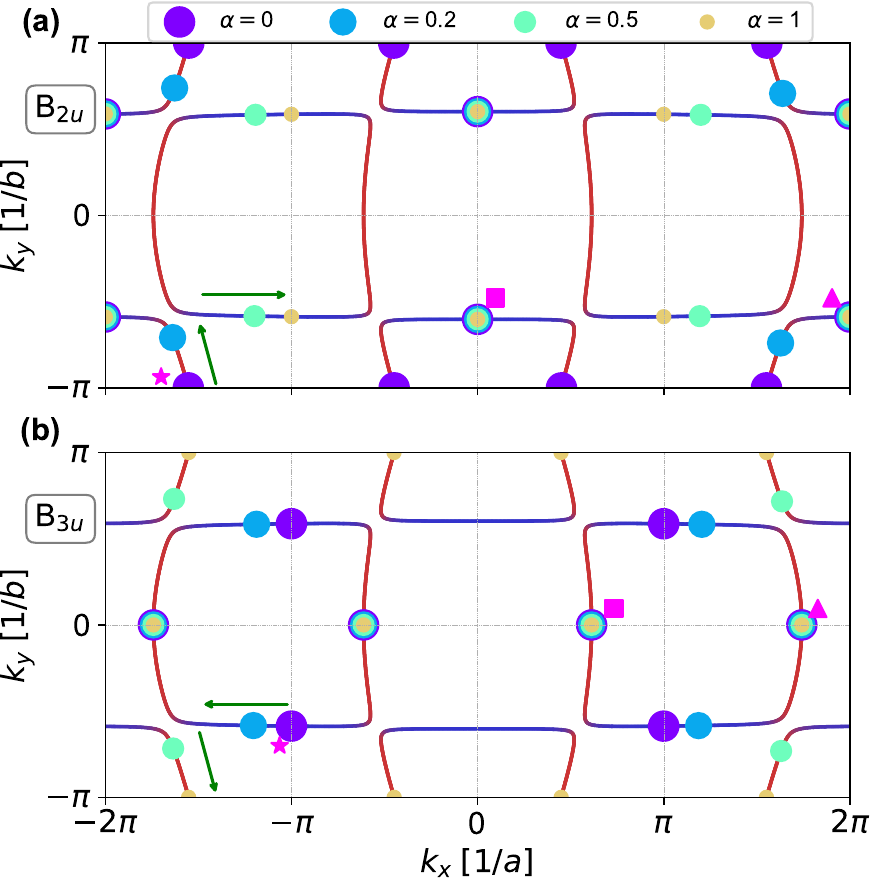}
    \caption{Overview of the positions of the symmetry-imposed and the additional nodes in the $k_z=0$ plane for B$_{2u}$ (a) and B$_{3u}$ (b) phases. In (a) relevant for B$_{2u}$, we display the gap function $d_{z,\mathrm{Te}},d_{z,\mathrm{U}}  \sim (1-\alpha)f_{x,1}+\alpha f_{x,2}$ for four different values of $\alpha=0,0.2,0.5,1$. Panel (b) is the same as (a) except for B$_{3u}$ with the corresponding relevant gap function  $d_{z,\mathrm{Te}},d_{z,\mathrm{U}}  \sim (1-\alpha)f_{y,1}+\alpha f_{y,2}$. As seen, the symmetry-imposed nodes (marked by a square and triangle) remain fixed, whereas the additional nodes (marked by a star) move with $\alpha$ as indicated by the green arrows in both panels.}
    \label{fig:movingpseduonodes}
\end{figure}

In Fig.~\ref{fig:movingpseduonodes}(a) we show the nodal structure of the B$_{2u}$ phase for different choices of coefficients. Specifically, for a $\vec d$-vector of the form $\vec d \sim (0,0,(1-\alpha)f_{x,1}+\alpha f_{x,2})$. This shows that any linear combination of pairing functions will change the locations of the additional nodes, but never gap them out. The same argument applies to the B$_{3u}$ phase shown in Fig.~\ref{fig:movingpseduonodes}(b). The robustness of the additional nodes is a result of the single component of the $\vec d$-vector, meaning that all superconducting terms necessarily commute, so the gap is just a superposition of the different pairing functions. Similarly, higher-order momentum functions as well as pairing in the U-Te hybridization sector will lead to contributions in the gap that share the same nodal lines as $f_{x,1}$ and $f_{x,2}$, and can therefore not gap out the additional nodes. This is a main result of the present paper, and implications of these additional nodes will be discussed further below.

\subsection{Superconducting topological surface states}

As a result of the odd parity of the superconducting order parameter, the boundaries of this topological system 
may exhibit zero-energy bound states in the form of Majorana-Dirac surface cones, that is, states with linear dispersion that form around time-reversal invariant momenta (TRIM) in the surface BZ \cite{buchholtz_1981,SatoFujimoto,Hsieh2012,Ishizuka,Geier_2020,Tei2023,Henrik}.

\begin{figure*}[tb]
    \centering
    \includegraphics[width=\linewidth]{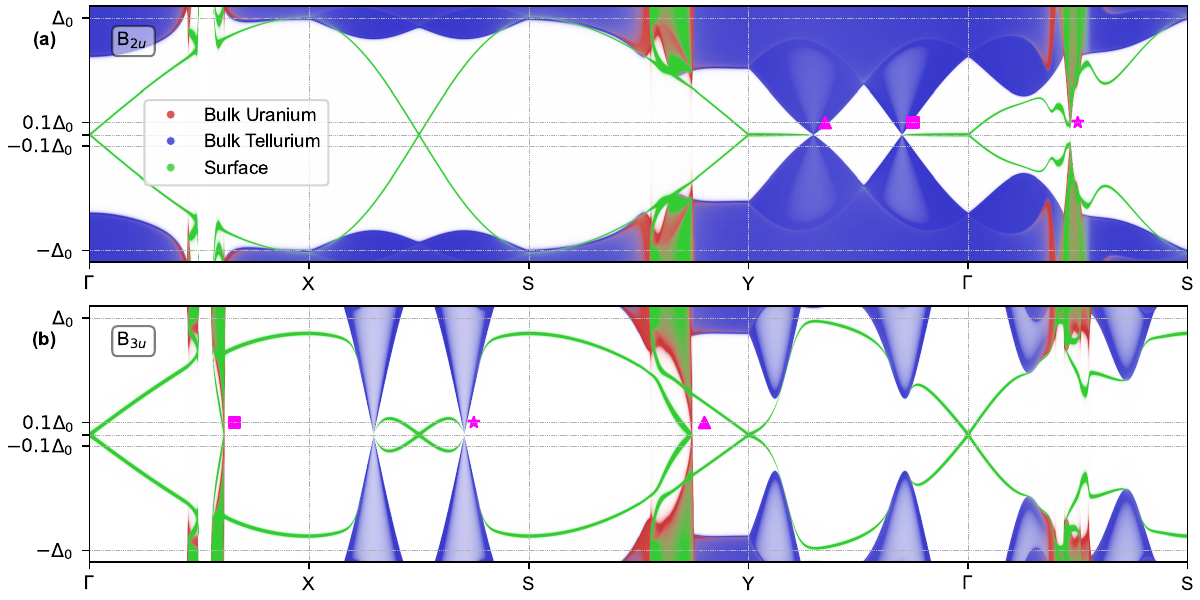}
    \caption{Spectral functions for the B$_{2u}$ (a) and B$_{3u}$ (b) phases while moving through the BZ path $\Gamma$XSY$\Gamma$S as defined in Fig.~\ref{fig:spectral_map}(b). Bulk bands are displayed in red (uranium) and blue (tellurium) colors, highlighting the nature of the nodes on the surface BZ. The green bands show the topological surface states on the (0-11) surface.}
    \label{fig:spectral_bzline}
\end{figure*}

\begin{figure}[tb]
    \centering
    \includegraphics[width=1.0\linewidth]{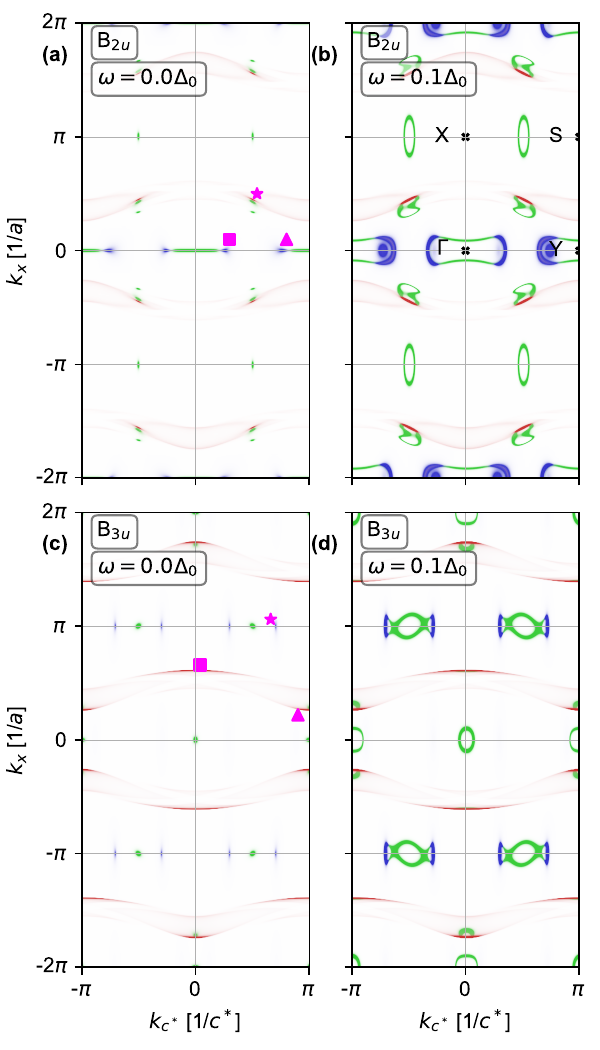}
    \caption{Bulk (red/blue) and surface (green) spectral functions for the B$_{2u}$ (a,b) and B$_{3u}$ (c,d) phases. Note that the $x$-axis is rotated from vertical to horizontal with respect to Fig.~\ref{fig:movingpseduonodes} and $k_{c^*}$ appears due to the (0-11) surface projection. The symbols in magenta indicate the locations of the symmetry-imposed nodes (square and triangle) as well as the additional nodes (star), consistent with the symbols plotted in Figs.~\ref{fig:movingpseduonodes} and \ref{fig:spectral_bzline}.}
    
    \label{fig:spectral_map}
\end{figure}

We will now consider in detail the states arising from the bulk B$_{2u}$ and B$_{3u}$ phases on the experimentally relevant (0-11) cleave plane. Using the method described in Sec.~\ref{sec:surface states}, we obtain the bulk and surface Green's functions from which we compute the bulk spectral densities of the uranium and tellurium sectors as well as the combined surface spectral density. The result for the choice of parameters with $\alpha=0$, according to the discussion in Sec.~\ref{sec:gapnodalstructure} is shown in Figs.~\ref{fig:spectral_bzline} and \ref{fig:spectral_map}. For this case of superconducting parameters, the additional nodes are located at $k_y=\pm \pi$ for B$_{2u}$ and at $k_x=\pm \pi$ for B$_{3u}$. On the (0-11) surface, this translates to additional nodes at $k_{c^*} = \pm \pi/2$ for B$_{2u}$ and $k_x = \pm \pi$ for B$_{3u}$ as indicated by the ($*$) symbols shown in Figs.~\ref{fig:movingpseduonodes}-\ref{fig:spectral_map}.

In order to gain insight into the topological nature of the surface bound states and determine whether there are any qualitative differences between the B$_{2u}$ versus B$_{3u}$ phases, we consider the symmetry representations for the two different phases. We will exploit that the (0-11) surface preserves only the $\mathcal M_x$ crystal symmetry along with time-reversal $\mathcal T$ and the particle-hole anti-symmetry $\mathcal P$. Since the characters of the B$_{2u}$ and B$_{3u}$ phases are opposite under $\mathcal M_x$, this leads to qualitative differences that are robust against microscopic changes of the model that preserve $\mathcal M_x$. We note that the even/odd property of B$_{3u}$/B$_{2u}$ under $\mathcal M_x$ implies that they do not mix on the (0-11) surface. Their distinct symmetries under $\mathcal M_x$ can be seen directly from Fig.~\ref{fig:2dfs_basisfunctions} with B$_{2u}$ (B$_{3u}$) being odd (even) under $\mathcal M_x$. Specifically, we will see that the B$_{2u}$ phase features Majorana flat bands around the $\Gamma$ point, while the B$_{3u}$ exhibits a Majorana-Dirac surface cone with linear dispersion in both $k_x$ and $k_{c^*}$.

First, we consider the algebra of the representations of the symmetries in the bulk. Following Ref.~\onlinecite{Geier_2020}, we express the commutation relation between the BdG representation $U$ of the crystal symmetry $g$ and the particle-hole anti-symmetry $\mathcal P$ in terms of the character $\chi(g)$ as
\begin{equation}
    U(g) U(\mathcal P) = \chi(g) U(\mathcal P) \mathcal U(g) ,\
\end{equation}
where the representation of particle-hole is given by $U(\mathcal P) = \tau_1 K$, where $\tau_1$ is the first Pauli matrix in particle-hole space and $K$ is complex conjugation. Due to the presence of strong SOC, the spin and momenta transform together, resulting in the normal state representation of the $\mathcal M_x$ symmetry given by $i\sigma_x$, where $\sigma_i$ are the Pauli matrices in spin-space. Both the bulk and the surface respect time-reversal symmetry and particle-hole anti-symmetry and therefore also the chiral anti-symmetry defined as $\mathcal C=\mathcal T \mathcal P$ \cite{Schnyder}. Note that chiral symmetry leaves the momentum invariant, acting only on the internal degrees of freedom. Since time-reversal symmetry $U(\mathcal T) = i\sigma_y K$ commutes with $\mathcal M_x$, the commutation relation between $\mathcal C$ and $\mathcal M_x$  are given by
\begin{equation}
    U(\mathcal M_x) U(\mathcal C) = \chi(\mathcal M_x) U(\mathcal C) U(\mathcal M_x),\label{eq:MxPcomm}
\end{equation}
where $\chi(\mathcal M_x)  = 1$ for B$_{2u}$ and $\chi(\mathcal M_x) =-1$ for B$_{3u}$ as seen from Tab.~\ref{tab:chractertable}. In the basis of the Majorana zero modes, a similar relation must hold for the representation of the symmetries of the surface, which in turn constrains the dispersion of the surface states. To see this, we write an effective surface Hamiltonian $H_s(\mathbf{k^\parallel})$ around the $\Gamma$ point. In both the B$_{2u}$ and B$_{3u}$ cases, we observe two localized surface states and we can therefore write it as a two-band model in terms of the Pauli matrices $\eta_i$. We denote the symmetry representation in the basis of the surface states as $\mathcal O(g)$, where $g=\mathcal M_x,\mathcal P,\mathcal T$ are the set of symmetry operations preserved by the (0-11) cleave plane. We define the basis such that time-reversal and particle-hole representations are given by 
\begin{align}
    & \mathcal O(\mathcal P) = \eta_0 K,\ \notag \\
    &\mathcal{O}(\mathcal T) = i\eta_2 K ,
\end{align}
such that $\mathcal O(\mathcal C) = i\eta_2$. 

This implies that the effective low-energy surface Hamiltonian can be written as 
\begin{equation}
    H_s(\mathbf{k^\parallel}) = f_1(\mathbf{k^\parallel}) \eta_1 + f_2(\mathbf{k^\parallel}) \eta_3,
\end{equation}
where both $f_1( \mathbf k^\parallel)$ and $f_2(\mathbf{k}^\parallel)$ must be odd under $\mathbf{k}^\parallel \mapsto - \mathbf{k^\parallel}$ due to the $\cal T$ and $\cal P$ symmetries. In the following, we use $\mathbf k^\parallel=(k_{c^*},k_x)$ for the in-plane momenta on the (0-11) surface. For the B$_{2u}$ phase, the commutation relation in Eq.~(\ref{eq:MxPcomm}) require the representation of $\mathcal{M}_x$ in the two-band low-energy subspace to be $\mathcal O(\mathcal M_x) = i\eta_2$, which further requires $f_1(\mathbf k^\parallel)$ and $f_2(\mathbf k^\parallel)$ to be odd in $k_x$ and even in $k_{c^*}$. 
Thus, for the B$_{2u}$ phase, this implies a flat dispersion in $k_{c^*}$ around the TRIM that host these anomalous surface states. This is consistent with the numerical results shown in Figs.~\ref{fig:spectral_bzline} and \ref{fig:spectral_map} where they appear around the $\Gamma$ and $Y$ points in the surface BZ.  

In the B$_{3u}$ phase, where Eq.~(\ref{eq:MxPcomm}) now implies $\mathcal O(\mathcal M_x) = i\eta_1$ (or $\mathcal O(\mathcal M_x) = i\eta_3$), so that $f_1(\mathbf{k^\parallel)}$ is even and $f_2(\mathbf{k^\parallel)}$ is odd in $k_x$ (or $f_1(\mathbf{k^\parallel)}$ odd and $f_2(\mathbf{k^\parallel)}$ even in $k_x$), which is consistent with the presence of a Majorana-Dirac surface cone centered at the $\Gamma$ point, see Figs.~\ref{fig:spectral_bzline} and \ref{fig:spectral_map}. The same logic applies to any of the TRIM points shown in Fig.~\ref{fig:surfacebz} that are invariant under $\mathcal M_x$.

The above argument relies on the fact that only one crystal symmetry remains on the surface, under which B$_{3u}$ and B$_{2u}$ transform differently. This should be contrasted to a related discussion in Ref.~\onlinecite{Tei2023} where the authors calculate the surface states on the (001) surface. That is, a surface preserving both $\mathcal M_x$ and $\mathcal M_y$ symmetries, such that no qualitative difference can be made between the B$_{2u}$ and B$_{3u}$ phases. Indeed, they find that both phases generically host flat bands around the $\Gamma$ point~\cite{Tei2023}. 

\section{Discussion and Conclusions}\label{sec:conclusions}

In this paper we have started from the basic assumption that the low-energy electronic bands of UTe$_2$ consist of 2D sheets in agreement with recent quantum oscillation measurements~\cite{Eaton2024,Weinberger2024}, and captured by the tight-binding model described in Sec.~\ref{sec:methods} incorporating both U and Te orbital-dominated bands. We have focused on the properties of the B$_{2u}$ and B$_{3u}$ superconducting phases due to experimental evidence for nodal spin-triplet pairing structure. Such triplet phases may be stabilized by ferromagnetic Ising-like fluctuations~\cite{Shishidou_2021,IshizukaPRB2021,Tei2024}. We find from basic symmetry arguments that several nodes necessarily must be present in the bulk superconducting gap structure for both $B_{2u}$ and B$_{3u}$, in addition to the standard symmetry-imposed nodes along the high-symmetry axes. These additional nodes exist due to the extended shape of the Fermi surface of UTe$_2$. While it remains to be determined what are the precise observable consequences of these additional nodes in terms of transport and thermodynamics, we speculate that they may be relevant for the understanding of the temperature dependence of the penetration depth. Experimentally, all magnetic field directions exhibit close to $T^2$ behavior, which contradicts simple pairs of point nodes giving $T^4$ behavior when the applied magnetic field is directed along the point node direction~\cite{Ishihara2023}. For this reason, Ref.~\onlinecite{Ishihara2023} proposed a chiral superconducting ground state with B$_{3u}+i\mathrm{A}_{u}$ order where the point nodes are pushed away from the high-symmetry axes~\cite{Henrik}. In our theory, the high-symmetry point nodes remain, as required by pure B$_{2u}$ or B$_{3u}$ order, but the additional nodes may dominate the excited quasiparticle response leading to the observed close-to $T^2$ behavior of the penetration depth~\cite{Ishihara2023,Bae2021,Carlton-Jones,Iguchi2023}.

Spin-triplet superconductivity is prone to host topological surface states protected by bulk 3D winding numbers or crystalline symmetries, depending on the Fermi surface topology. Several previous theoretical studies have analyzed such topological Majorana edge states for UTe$_2$~\cite{Ishizuka,Shishidou_2021,Tei2023,Ohashi2024,Henrik}, and some experimental signatures have also been analyzed in terms of them~\cite{Jiao2020,Bae2021}. More specifically, in the case of time-reversal symmetric superconductivity and open cylindrical Fermi surfaces, which is topologically trivial in the sense of the AZ classification (3D class DIII~\cite{Schnyder}), Tei {\it et al.}~\cite{Tei2023} derived the Majorana edge states on high-symmetry surfaces perpendicular to the axes of the conventional unit cell, and identified the crystalline mirror and twofold rotational symmetries that protect them. In this paper, we have also analyzed the superconducting gap structure in the presence of cylindrical open Fermi surfaces. However, we have included both the U and Te bands and accessed the topological surface states on the experimentally relevant (0-11) cleave plane. This surface breaks all other symmetries except the mirror symmetry $\mathcal M_x$ (see Fig.~\ref{fig_elementary_cells}) under which the irreps B$_{2u}$ and B$_{3u}$ transform differently. Conveniently, this leads to qualitative differences in the dispersion of the surface states, offering a potential observable difference between these two phases in terms of surface-sensitive spectroscopic probes such as photoemission and scanning tunneling microscopy.

\begin{figure}[tb]
    \centering
    \includegraphics[width=\linewidth]{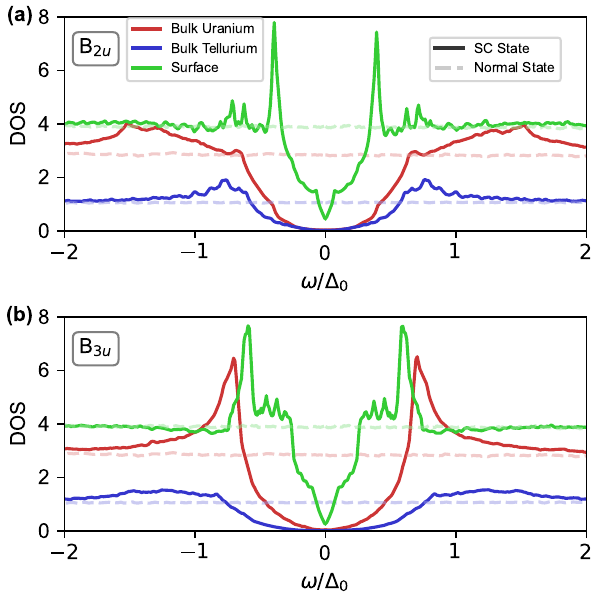}
    \caption{Density of states versus energy for both bulk and surface, obtained by ${N}(\omega) = \frac{-1}{\pi N_{\mathbf k}}\sum_{\mathbf{k}^\parallel}\Im \Tr [P_e G_{s/b}(\mathbf{k}^\parallel,\omega)]$ with $\Delta_0=10\: \mathrm{meV}$ and $N_k$ is the number of momentum points. Panel (a) and (b) correspond to B$_{2u}$ and B$_{3u}$ order parameters, respectively. Note that the normal state density of states is featureless in the entire energy range. Thus, smaller values of $\Delta_0$ yields identical results up to an overall scale.}
    \label{fig:dos}
\end{figure}

The low-energy surface states naturally give rise to "coherence peaks" below the actual bulk coherence peaks of the superconducting state, as seen from Fig.~\ref{fig:dos}. This points to a general feature of tunneling into spin-triplet superconductors: the extracted gap from surface sensitive probes may well yield smaller gaps than those extracted from bulk measurement probes. In fact, for UTe$_2$ this may be the case since the neutron resonance peak near $E_r\simeq 2\Delta\simeq1$ meV and an STM-extracted gap of $\Delta\simeq 0.25$ meV~\cite{GuArXiv} leads to an unusually large ratio $E_r/2\Delta\simeq 2$ compared to other superconductors also featuring a neutron resonance peak~\cite{Duan2021}. One may speculate that the bulk superconducting gap is larger than $\Delta\simeq 0.25$ meV, at least if STM indeed picks up tunneling contributions from topological surface states.

We note that a recent experimental STM quasiparticle interference (QPI) study reported enhanced scattering intensity in the superconducting phase at specific scattering wavevectors~\cite{SeamusQPI}. In an accompanying work, we have developed the theoretical framework for addressing QPI in spin-triplet superconductors and combined it with the current model for superconductivity in UTe$_2$ on the (0-11) surface~\cite{HansQPI}. A main conclusion of that study is that B$_{3u}$ appears more consistent with the currently available QPI data. We refer to Ref.~\onlinecite{HansQPI} for further details of the theoretical QPI analysis relevant for UTe$_2$.

We have departed the theoretical discussion in this paper based on the open cylindrical Fermi surface reported from recent quantum oscillation measurements~\cite{Eaton2024,Weinberger2024}. The existence of a small closed Fermi pocket, for example near $\Gamma$ or $Z$, as discussed in some works~\cite{Fujimori2019,Miao2020,Broyles2023,Choi2024}, alters some of the conclusions extracted from our analysis. For example, it opens up the possibility of B$_{1u}$ being nodal and changes the topological classification of the surface states as analyzed in several earlier works~\cite{Ishizuka,Shishidou_2021,Henrik}. However, the existence of a small Fermi pocket will not change our main conclusions about the properties and distinctions between pairing symmetries B$_{2u}$ and B$_{3u}$. In summary, the main conclusions of the present study include: 1)  the inevitable existence of nodes in addition to the symmetry-imposed nodes along high-symmetry axes due to the shape of the Fermi surface in UTe$_2$, 2) the existence of characteristic low-energy surface states emanating from the TRIM points of the experimentally relevant (0-11) cleave plane, and 3) distinct topological surface states between the B$_{2u}$ and B$_{3u}$ phases due to their opposite behavior under a single remaining mirror symmetry on that plane. These findings open up new possible explorations of further experimental consequences of the additional nodes and the surface states to help pinpoint the preferred pairing symmetry of the fascinating material UTe$_2$.

\begin{acknowledgments}
We acknowledge fruitful discussions with Daniel Agterberg, J.~C.~Séamus Davis, Peter J. Hirschfeld, and Shuqui Wang. H.~C. acknow\-ledges support from the Novo Nordisk Foundation grant NNF20OC0060019. A.K. acknowledges support by the Danish National Committee for Research Infrastructure (NUFI) through the ESS-Lighthouse Q-MAT.
\end{acknowledgments}

\appendix
\renewcommand{\thefigure}{S\arabic{figure}}
\setcounter{figure}{0}

\section{(0-11) Lattice transformation}\label{app:A}

\subsection{Primitive lattice}
\begin{figure}
    \centering
    \includegraphics[width=0.95\linewidth]{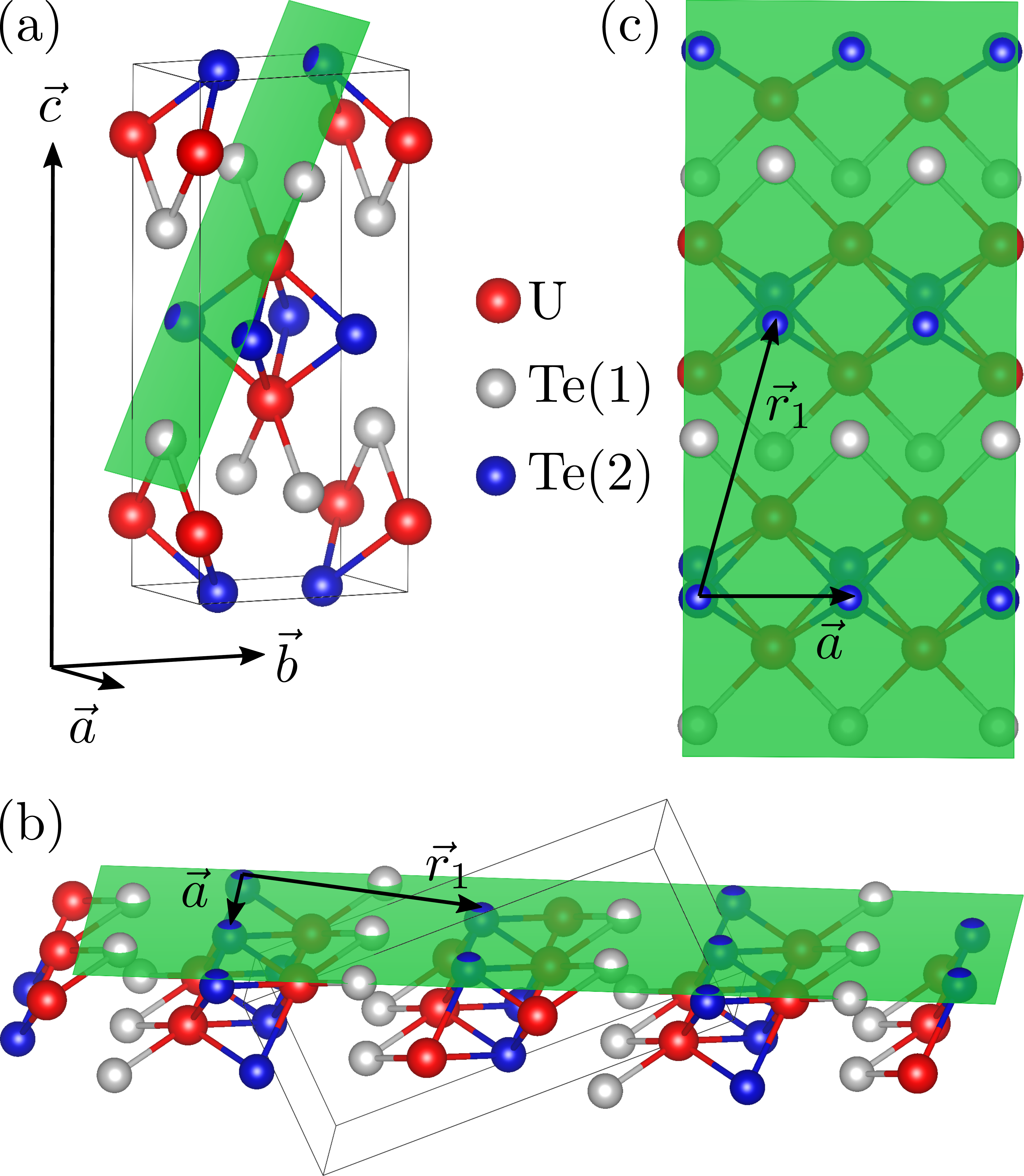}
    \caption{Crystal structure and cleave plane of UTe$_2$: (a) Conventional elementary cell together with cleave plane (green). (b) First layer together with lattice vectors $\vec a$ and $\vec r_1$ on the surface. (c) View down on the surface which exhibits mirror symmetry $\mathcal M_x$, but not in the othogonal direction because of the (white) Te atoms.}
    \label{fig_elementary_cells}
\end{figure}

Given the elementary cell of UTe$_2$ as shown in Fig. \ref{fig_elementary_cells}, the primitive lattice vectors can be written in the Cartesian coordinate system as
\begin{equation}
    R_{\mathrm{prim}} \equiv \begin{pmatrix}
        \Vec{r}_1 & \Vec{r}_2 & \Vec{r}_3
    \end{pmatrix}
    = \begin{pmatrix}
        -a/2 & a/2 & a/2 \\
        b/2 & -b/2 & b/2 \\
        c/2 & c/2 & -c/2
    \end{pmatrix},
\end{equation}
such that the primitive reciprocal lattice vectors become
\begin{equation}
    M_{\mathrm{prim}} \equiv 
    \begin{pmatrix}
        \Vec{m}_1 & \Vec{m}_2 & \Vec{m}_3 
    \end{pmatrix}^T
    = 
    \begin{pmatrix}
        0 & 1/a & 1/a \\
        1/b & 0 & 1/b \\
        1/c & 1/c & 0
    \end{pmatrix}^T .\
\end{equation}
To obtain the transformation from the components $k_i$ ($i=1,2,3$) of the lattice vectors in the primitive setting
to the ones in the conventional setting $i=x,y,z$ we set
\begin{equation}
    \begin{pmatrix}
        k_1 \\ k_2 \\ k_3 
    \end{pmatrix}    
    = M_{\mathrm{cart}}^{-1} M_{\mathrm{prim}} 
    \begin{pmatrix}
    k_x \\ k_y \\ k_z        
    \end{pmatrix} ,\
\end{equation}
where $M_{\mathrm{cart}}=\mathrm{diag}(1/a,1/b,1/c)$ is the matrix of Cartesian reciprocal vector.
This yields the two transformations
\begin{align}\label{primtocart}
    \begin{pmatrix}
        k_x \\ k_y \\ k_z
    \end{pmatrix}
    =
    \begin{pmatrix}
        k_2 + k_3 \\
        k_3 + k_1 \\
        k_1 + k_2
    \end{pmatrix} ,\
    &&
    \begin{pmatrix}
        k_1 \\ k_2 \\ k_3
    \end{pmatrix}
    =
    \frac{1}{2}
    \begin{pmatrix}
        -k_x + k_y + k_z \\
        k_x - k_y + k_z \\
        k_x + k_y - k_z
    \end{pmatrix} .\
\end{align}

It is convenient to use that 
\begin{align}
    & \mathrm{trig}^{n_1}\frac{k_x}{2}\mathrm{trig}^{n_2}\frac{k_y}{2}\mathrm{trig}^{n_3}\frac{k_z}{2} = \notag \\
    &\frac{1}{4}\qty(\mathrm{trig}^{\sum_j n_j}\qty(k_1+k_2+k_3) + \sum_i (-1)^{n_i}\mathrm{trig}^{\sum_j n_j}k_i),
\end{align}
where $\mathrm{trig}^{n}k = \cos k,\sin k, -\cos k, -\sin k$ for $n=0,1,2,3$ respectively, and $n$ should be evaluated modulo 4.
This shows that the normal-state Hamiltonian can be written in terms of functions of integer multiples of the primitive momentum coordinates only, which means that we can obtain the corresponding real-space tight-binding model simply by replacing each Fourier coefficient $\mathrm{e}^{-ik_i}$ with its corresponding real-space translation operator $T_{\vec r_i}$.

\subsection{Transformation to (0-11) coordinate system}

The surface accessible in experiments is the one with normal vector
\begin{equation}
    \hat n = \frac{\Vec{a}\cross (\vec{b}+\Vec{c})}{\abs{\Vec a}\abs{\Vec c + \vec b}} = \frac{-c\hat e_y + b \hat e_z}{\sqrt{b^2+c^2}} ,\
\end{equation}
where we use $\Vec{a}=a\hat e_x$, $\Vec{b}=b\hat{e}_y$ and $\Vec{c} = c\hat e_z$ and $a = 4.161 \: \mathrm{Å}$, $b=6.122 \: \mathrm{Å}$ and $c=13.955 \: \mathrm{Å}$ are the lattice constant along the three crystal axes \cite{Jiao2020}. The angle between $\hat e_y$ and the vector normal to the surface is given by $\cos \theta = c/\sqrt{b^2+c^2}$ with $\theta = 23.7^{\circ}$.

To properly capture the physics at the (0-11) surface, we first need to transform to a coordinate system that respects the symmetry of the surface, that is, we need two lattice vectors that are orthogonal to $\hat n$ and an integer multiple of the primitive lattice vectors. It is convenient to choose these as $\vec r_1$ and $\vec r_+ = \vec r_2 + \vec r_3$ and we choose the last vector of the coordinate system as $\vec r_3$. We can now perform a Fourier transformation in the parallel momenta to obtain a Hamiltonian in the form 
\begin{equation} \label{eq:hamiltontridiagonal}
    H(\mathbf{k}^\parallel) = H_0(\mathbf{k}^{\parallel} ) + T_{\vec r_3}V(\mathbf{k}^\parallel) + T_{\vec r_3}^T V^\dag(\mathbf{k}^\parallel) ,\
\end{equation}
where $T_{\vec r_3}$ is the real space operator that translates by $\vec r_3$. From this form, we can use the method described in Ref.~\onlinecite{sancho_highly_1985} to obtain the Green's function components on the surface and bulk of the system with respect to the $\vec r_3$ direction. That is, we obtain a triple of Green's functions $(G_s(\omega,\mathbf{k}^\parallel),G_b(\omega,\mathbf{k}^\parallel),\tilde G_{{s}}(\omega,\mathbf{k}^\parallel))$ as a function of energy $\omega$ and parallel momentum
\begin{equation}
     \mathbf{k}^\parallel = k_1 \vec m_1 + k_+ \vec m_+  .\
\end{equation}
Finally, we choose the coordinate system with $\vec m_{a^*} = \vec m_+ - \vec m_1/2$ and $\vec m_{c^*} = \vec m_1$, and correspondingly $k_{a^*} = k_+$ and $k_{c^*} = k_1 + k_+/2$ so we can write 
\begin{equation}\label{eq:newkparallel}
    \mathbf{k}^\parallel = k_{a^*} \vec m_{a^*} + k_{c^*} \vec m_{c^*},
\end{equation}
where $\vec m_{a^*}$ and $\vec m_{c^*}$ are orthogonal. We identify $k_{a^*}=k_x$ since $\vec m_{a^*} = \hat e_x/a$ while $\vec{m}_c^* = \hat e_y/b + \hat e_z/c$. In the coordinates defined by Eq.~(\ref{eq:newkparallel}), the reciprocal lattice vectors are given by $(2\pi,0)$ and $(\pi,2\pi)$. 

Note that in the figures in the main text, we reinstate the units of the momentum vectors, such that $k_x$ is measured in units of $1/a$ and $k_{c^*}$ in units of $1/c^*$ with $c^* = \sqrt{c^2+b^2}/2$·

\section{Bulk and surface Green's functions.}\label{app:B}

We consider a class of Hamiltonians acting on the Hilbert space 
\begin{equation}
    \mathcal H = \bigoplus_{n=0}^N \mathcal H_n ,\
\end{equation}
with basis vectors $\ket{n,\alpha}$ where $\alpha$ labels a complete basis on the $\mathcal H_n$ subspaces. The Hamiltonian is represented as 
\begin{equation}
    H = \mathds 1 \otimes H_0 + T \otimes V + T^T \otimes V^\dag,
\end{equation}
where $T$ is the translation operator acting as 
\begin{equation}
    T \ket{n,\alpha} = \ket{n-1,\alpha} ,\
\end{equation}
and $\mathds 1$ denotes the identity. We can also write it in its matrix block form as

\begin{equation}\label{eq:tridiag}
    H =     \begin{pmatrix}
    H_0 & V  &  &  &  \\
    V^\dag & H_0 & V &  &   \\
    & V^\dag & H_0 & \ddots &   \\
     & & \ddots & \ddots & V  \\
    & & & V^\dag & H_0
    \end{pmatrix} .\
\end{equation}

Note that this system exhibits translational invariance except at the boundaries of the system. Note also that it is possible to bring any system that has translation symmetry except at the boundary into this form simply by extending the size of the unit cell until each cell is connected only through nearest-neighbor interactions. We now wish to derive an iterative procedure to find the Green's function in the first block of the system, as described in Ref.~\onlinecite{sancho_highly_1985}.

Let $G$ be the Green's function of the system defined by 
\begin{equation}\label{eq:Gdef}
    (\omega-H)G = \mathds 1 .
\end{equation}
For this to be well defined at the energy levels of $H$ we use $\omega \mapsto \omega+i\eta$ for an infinitesimal $\eta$, which moves the poles of $G(\omega)$ to the upper half of the complex plane \cite{bruusflensberg2003}. The blocks of $G$ are defined as 
\begin{equation}
    G_{nn'} = P_n G P_{n'},
\end{equation}
where $P_n$ is the projector onto the $n$'th layer subspace, given by 
\begin{equation}
    P_n = \sum_\alpha \ketbra{n,\alpha}{n,\alpha} .\
\end{equation}
Expanding Eq.~(\ref{eq:Gdef}) into the blocks from Eq.~(\ref{eq:tridiag}), we obtain the system of equations 
\begin{equation}\label{eq:0eq}
    (\omega-H_{0})G_{00} = \mathds 1 + VG_{10} ,\
\end{equation}
and
\begin{equation}\label{eq:neq}
    (\omega-H_{0})G_{n0} = V^\dag G_{n-1,0}+VG_{n+1,0} .\
\end{equation}

Inserting Eq.~(\ref{eq:neq}) for $n=1$ into Eq.~(\ref{eq:0eq}), we obtain
\begin{equation}
    (\omega - H_0 - V (\omega-H_0)^{-1}V^\dag)G_{00} = \mathds 1 + V (\omega-H_0)^{-1} V G_{20},
\end{equation}
which is isomorphic to Eq.~(\ref{eq:0eq}) with renormalized coefficients and $G_{10}\mapsto G_{20}$. Similarly, we can write a renormalized version of Eq.~(\ref{eq:neq}) as 
\begin{align}
    &\left(\omega-H_0-V(\omega-H_0)^{-1}V^\dag - V^\dag(\omega-H_0)^{-1}V\right)G_{n0}, \\
    &= V^\dag(\omega-H_0)^{-1}V^\dag G_{n-2,0}+V(\omega-H_0)^{-1}VG_{n+2,0}, 
\end{align}
where we have $G_{n\pm 1,0}\mapsto G_{n\pm 2,0}$ and renormalized coefficients. Thus, we see that the effect of this renormalization is to acquire equations connecting the next-nearest layers. One can repeat this procedure $l$ times to obtain the set of equations 
\begin{equation}
    (\omega-\varepsilon_l^s)G_{00} = \mathds 1 + \alpha_l G_{2^ln, 0},
\end{equation}
and 
\begin{equation}
    (\omega-\varepsilon_l)G_{2^ln,0} = \beta_l G_{2^l(n-1),0}+\alpha_l, G_{2^l(n+1),0}
\end{equation}
where the coefficients can be obtained through the iterative relations 
\begin{equation}
    \alpha_i = \alpha_{i-1}(\omega-\varepsilon_{i-1})^{-1}\alpha_{i-1} ,\
\end{equation}
\begin{equation}
    \beta_i = \beta_{i-1}(\omega-\varepsilon_{i-1})^{-1}\beta_{i-1} ,\
\end{equation}
\begin{equation}
    \varepsilon_i = \varepsilon_{i-1} + \alpha_{i-1}(\omega-\varepsilon_{i-1})^{-1}\beta_{i-1}+\beta_{i-1}(\omega-\varepsilon_{i-1})^{-1}\alpha_{i-1} ,\
\end{equation}
\begin{equation}
    \varepsilon_{i}^s = \varepsilon_{i-1}^s+\alpha_{i-1}(\omega-\varepsilon_{i-1})^{-1}\beta_{i-1} ,\
\end{equation}
with the initial conditions $\varepsilon_0 = \varepsilon^s_0 = H_0$, $\alpha_0 = V$ and $\beta_0 = V^\dag$. Now for sufficiently large $l$, such that $2^l$ is much larger than the range of correlations, the coefficients connecting $G_{2^ln,0}$ with $G_{2^l(n-1),0}$ and $G_{2^l(n+1),0}$ must go to zero, that is $\alpha_l,\beta_l \to 0$. From the recursion relations we see that this implies that the sequences $\{\varepsilon_l\}$ and $\{\varepsilon_l^s\}$ are convergent. We will denote their limits by $M_b=\lim_{l\to\infty}\varepsilon_l$ and $M_s=\lim_{l\to\infty}\varepsilon_l^s$, respectively.  We then obtain the simple equation for the surface component of the Green's function
\begin{equation}\label{eq:Gseq}
    (\omega-M_s)G_{00} = \mathds 1,
\end{equation}
from which one can solve for the surface Green's function $G_s\equiv G_{00}$. Similarly, we also have
\begin{equation}
    (\omega-M_b)G_{2^l 2^l} = \mathds 1 .\
\end{equation}
For large $l$,  $G_{2^l2^l}\to G_b$ is the Green's function component describing a site far into the bulk of the system~\cite{sancho_highly_1985}.  


%

\end{document}